\documentclass[12pt]{book}

\usepackage[dvips]{graphicx,color}
\usepackage{makeidx,tsukuba}
\makeauthorindex
\makeindex

\begin{document}

\BookTitle{\itshape The 28th International Cosmic Ray Conference}
\CopyRight{\copyright 2003 by Universal Academy Press, Inc.}
%\tableofcontents
\pagenumbering{arabic}

\chapter{%   %%%%%%%%% <===== TITLE of the contribution
%%%%%%%%%%% The first letter of each word should be capital letter.
Measurement of Cosmic-Ray Proton, Antiproton and Muon Spectra 
at Mountain Altitude}

\author{%
%
% You can include as many co-authors as you wish, unless
% the title/author information fits within 1 page.
%
T.~Sanuki,$^1$ M.~Fujikawa,$^1$ K.~Abe,$^2$ K.~Anraku,$^{1,a}$ 
H.~Fuke,$^1$ S.~Haino,$^1$ M.~Imori,$^1$ K.~Izumi,$^1$ 
T.~Maeno,$^{2,b}$ Y.~Makida,$^3$ N.~Matsui,$^1$ H.~Matsumoto,$^1$ 
H.~Matsunaga,$^{1,c}$ J.~Nishimura,$^1$ M.~Nozaki,$^2$ 
S.~Orito,$^{1,*}$ M.~Sasaki,$^{3,d}$ Y.~Shikaze,$^2$ 
J.~Suzuki,$^3$ K.~Tanaka,$^3$ A.~Yamamoto,$^3$ Y.~Yamamoto,$^1$ 
K.~Yamato,$^2$ T.~Yoshida,$^3$ and K.~Yoshimura$^3$ \\
{\it (1) The University of Tokyo, Bunkyo, Tokyo 113-0033, Japan\\
(2) Kobe University, Kobe, Hyogo 657-8501, Japan\\
(3) High Energy Accelerator Research Organization (KEK), Tsukuba, Ibaraki 305-0801, Japan}
}%% end of author

\section*{Abstract}
Measurement of cosmic-ray proton, antiproton and muon spectra was carried out
 at mountain altitude. 
We observed $2 \times 10^5$ protons and $10^2$ antiprotons
 in a kinetic energy region of 0.25 -- 3.3~GeV. 
Zenith-angle dependence of proton fluxes was obtained. 
Atmospheric muon spectra 
 were measured simultaneously. 
The observed antiproton spectrum showed some deviation
 from theoretical predictions particularly in a low energy region. 

\section{Introduction}

Primary cosmic rays hit the Earth's atmosphere and produce
 baryons and mesons via hadronic interactions.
Absolute fluxes of these 
 ``secondary cosmic rays''
 can be calculated by using 
 the primary cosmic-ray intensity and interaction cross sections.
Observation of the secondary cosmic rays is very important
 to verify, or to improve, theoretical calculations.
It is essentially important to understand
 propagation process of the secondary particles inside the atmosphere.
We report new measurement of
 secondary cosmic ray spectra
 at  mountain altitude. 

\section{Observations}

We performed cosmic-ray observation 
 at Norikura Observatory, ICRR, University of Tokyo, Japan,
 in September 1999,
%% with the BESS detector \cite{detector}\cite{agel}\cite{orito}\cite{newtof},
 with the BESS detector [1,2,11,17],
 which was the same apparatus as we utilized to measure
 the primary protons and antiprotons
%% \cite{pbasaoka}\cite{pbmaeno}\cite{pborito}\cite{bib:bessphe},
 [3,9,12,13],
%% as well as atmospheric muons at sea level \cite{motokimu}. 
 as well as atmospheric muons at sea level [10]. 
The observatory is located at 2,770m above sea level. 
%%The vertical cutoff rigidities is 11.2~GV \cite{shea}.
The vertical cutoff rigidities is 11.2~GV [16].
During the observation,
 the mean atmospheric depth was
 742~${\mathrm {g/cm^2}}$. 

\section{Results and Discussion}

The measured energy spectra of protons and antiprotons are shown
 in 
%% Fig.~\ref{fig:antiproton-flux},
 Fig.\ 1,
 together with the previous measurements at mountain altitude
%% \cite{barber80}\cite{koch54}\cite{koch58}\cite{sembroski86}. 
 [4,7,8,15].
The antiproton spectrum is compared
%% with theoretical predictions \cite{bowen86}\cite{Stephens96}. 
 with theoretical predictions [5,18]. 
In this analysis, the zenith angle $(\theta _z)$ was limited as
 $\cos \theta _z \geq 0.95$ for protons, and
 $\cos \theta _z \geq 0.84$ for antiprotons,
 thus the obtained fluxes are ``near-vertical'' fluxes. 
\begin{figure}
  \begin{center}
    \includegraphics[width=7.2cm]{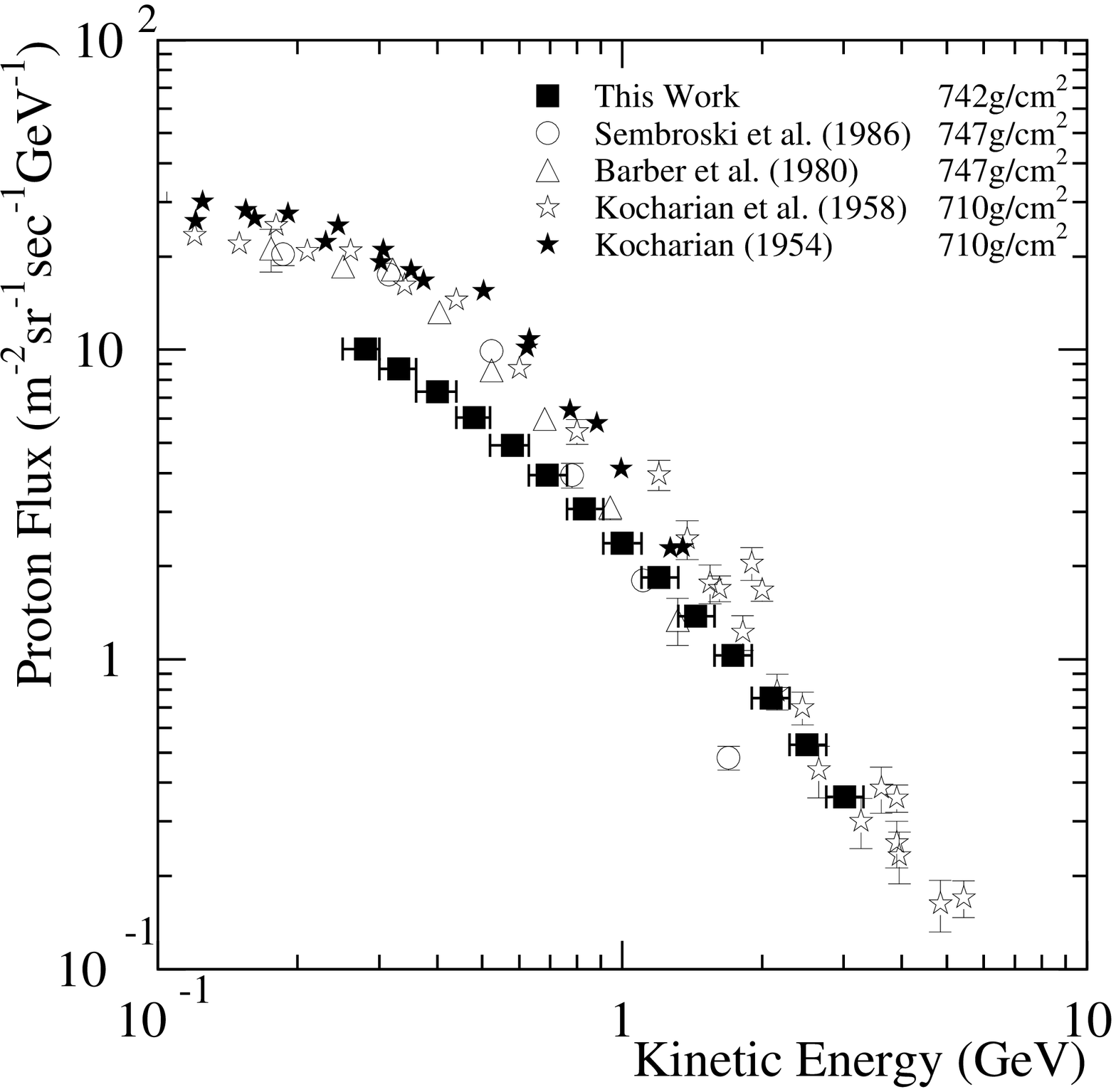}
    \includegraphics[width=7.2cm]{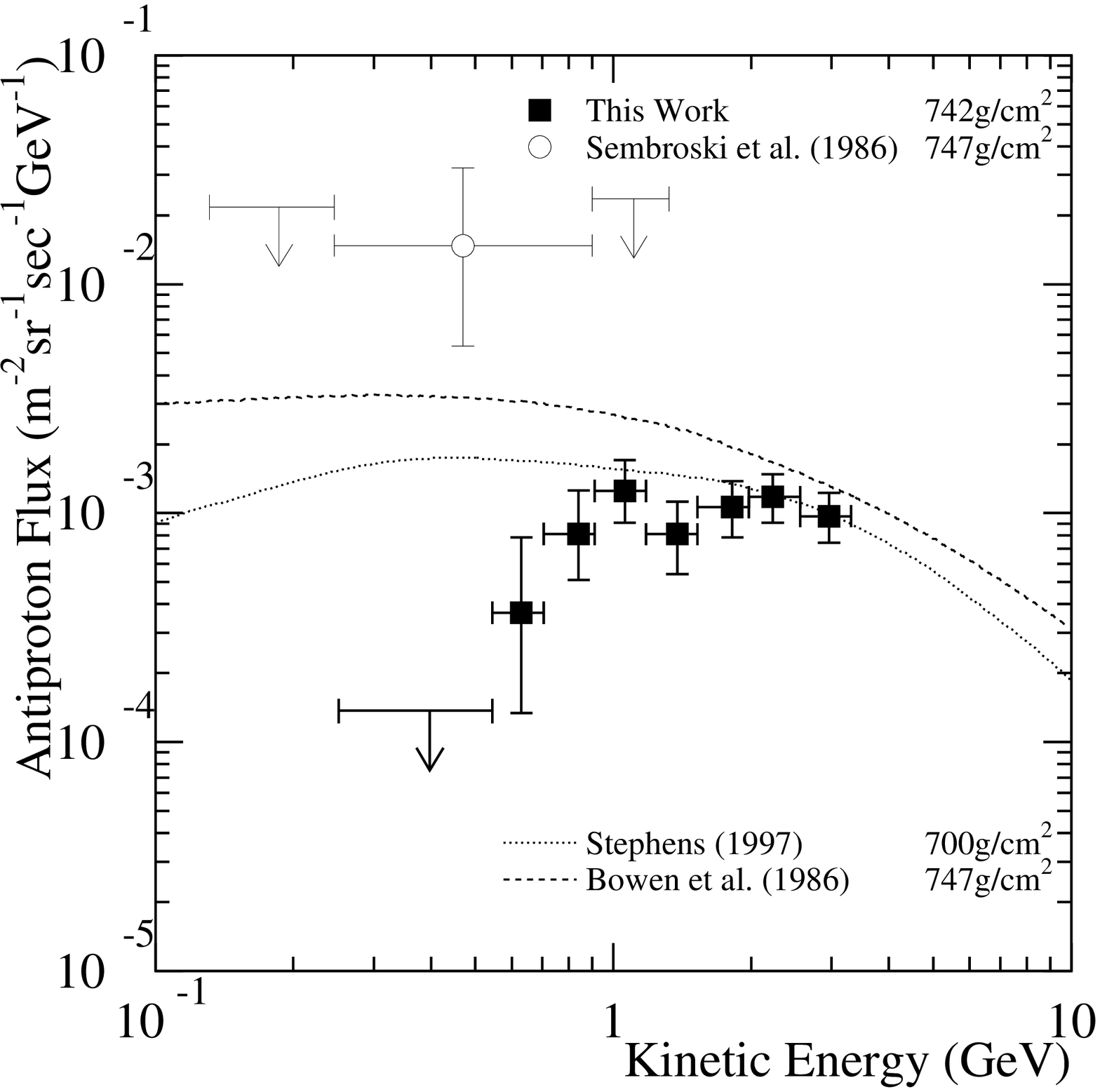}
  \end{center} 
  \vspace{-0.5pc}
  \caption{\label{fig:antiproton-flux}
  Observed near-vertical flux of protons (left) and antiprotons (right). 
  Calculated antiproton spectra are compared with the observed ones.}
\end{figure}
There is some disagreement among the proton fluxes
%% shown in Fig.~\ref{fig:antiproton-flux}. 
 shown in Fig.\ 1. 
According to simple Monte Carlo simulations,
 the deviations can be explained
 by the different altitudes and cutoff rigidities at their observation sites.

%%Fig.~\ref{fig:zenith-angle-proton-flux} shows
Fig.\ 2 shows
 zenith angle dependence of the observed proton flux
 in two kinetic energy regions. 
\begin{figure}
  \begin{center}
    \includegraphics[width=8.7cm]{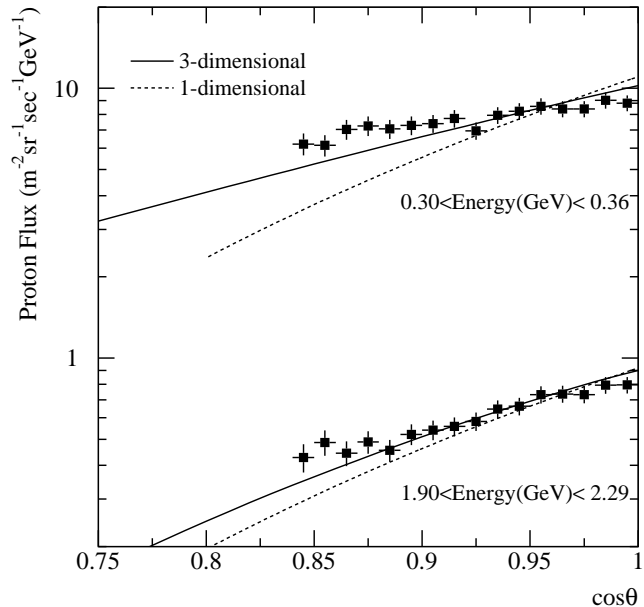}
  \end{center}
  \vspace{-0.5pc}
  \caption{\label{fig:zenith-angle-proton-flux}
  Zenith angle dependence of proton flux. 
  The dotted and solid lines show the expected dependence
  in simple one-dimensional and three-dimensional calculations,
  respectively.
   }
\end{figure}
The zenith angle dependence can be expected
 in a simple one-dimensional approximation,
 $F(\cos\theta _z)=F_{0}\exp(X/\lambda (1-1/\cos\theta _z))$,
 where $X$ is the atmospheric depth 
 and $\lambda$ is the absorption mean free path of protons
 inside the atmosphere. 
%%Dotted lines in Fig.~\ref{fig:zenith-angle-proton-flux} show the expectation,
Dotted lines in Fig.\ 2 show the expectation,
 in which $X/\lambda=6$ was assumed. 
Relatively good agreement was found 
 between the observed data and the calculation
 in the higher energy region. 
In the lower energy region, however,
 significant discrepancy was found between them. 
These facts are
 most likely due to the effect of angular spread of secondary protons
 produced via nuclear interactions. 
%%The solid lines shown in Fig.~\ref{fig:zenith-angle-proton-flux} give
The solid lines shown in Fig.\ 2 give
 the results of the analytic calculation,
 in which the angular spread was taken into account.
They reproduced the observed data better than one-dimensional calculation
 in the whole energy range. 

The observed antiproton spectrum agrees
%% with a theoretical calculation by Stephens \cite{Stephens96} above 1~GeV
 with a theoretical calculation by Stephens [18] above 1~GeV
%% as shown in Fig.~\ref{fig:antiproton-flux}. 
 as shown in Fig.\ 1. 
On the other hand, the flux below 1~GeV shows significant disagreement. 
%%In Ref.~\cite{Stephens96},
In Ref.~[18],
 production spectra of antiprotons
 are calculated,
 and the result shows a sharp peak around 2~GeV. 
This means that most antiprotons observed below 1~GeV are
 tertiary antiprotons, 
 those which have been produced inside the atmosphere
 and then lost their energies
 during the propagations in the atmosphere. 
In this case, cross sections in $\bar{p} + A (nuclei)$ processes are
 to be precisely treated
 for an accurate evaluation of antiproton spectrum at mountain altitude. 
A recent work made with Monte Carlo simulation reported a preliminary result
%% which shows better agreement with the observed spectrum \cite{buenerd}. 
 which shows better agreement with the observed spectrum [6]. 

The atmospheric muon spectra
 in a momentum range of 0.6 -- 106~GeV/$c$
 have been measured simultaneously. 
%%The results are reported in Ref. \cite{bib:norimu}. 
The results are reported in Ref.~[14]. 
The observed muon spectra showed much better agreement
 with theoretical prediction than that of antiprotons. 
It shows that secondary meson productions are treated rather properly
 in the theoretical calculations. 
However, 
 secondary baryon interaction cross sections
 have to be modified
 so as to reproduce the observed antiproton spectrum. 

\section{Summary}

We have measured proton, antiproton and muon spectra
 at Mt. Norikura, Japan,
 where the atmospheric depth was 742~${\mathrm {g/cm^2}}$. 
The zenith angle dependence in the proton flux was observed. 
It suggests an importance of three-dimensional effect
 of angular spread in secondary baryon productions. 
The calculated antiproton flux agrees
 with our antiproton flux above 1~GeV.
In a lower energy region,
 however,
 our measurement gives much lower flux than that of the prediction.

\section*{Acknowledgment}

This study was supported by the Joint Research Program of
 ICRR, the University of Tokyo and
 Grants-in-Aid, KAKENHI(11694104, 11440085, 09304033), from MEXT and JSPS.
We would like to thank NASA, ISAS, KEK and ICEPP, the University of Tokyo
 for their continuous support. 

\vspace{\baselineskip}
\re
$^a$ Present address: Kanagawa University, Yokohama, Kanagawa 221-8686, Japan
\re
$^b$ Present address: CERN, CH-1211 Geneva 23, Switzerland
\re
$^c$ Present address: University of Tsukuba, Tsukuba, Ibaraki 305-8571, Japan
\re
$^d$ Present address: NASA/GSFC, Greenbelt, MD 20771, USA
\re
$^*$ deceased.

\vspace{\baselineskip}
\re
1.\ Ajima Y. et al. \ 2000, Nucl. Instr. and Meth. A 443, 71
\re
2.\ Asaoka Y. et al.\ 1998, Nucl. Instr. and Meth. A 416, 236
\re
3.\ Asaoka Y. et al.\ 2002, Phys. Rev. Lett. 88, 051101
\re
4.\ Barber H. B. et al.\ 1980, Phys. Rev. D 22, 2667
\re
5.\ Bowen T. and Moats A.\ 1986, Phys. Rev. D 22, 2667
\re
6.\ Bu\'{e}nerd M.\ 2002, Int. J. Mod. Phys. A 17, 1665
\re
7.\ Kocharian N. M. \ 1954, J. Exper. Theoret. Phys. (USSR) 28, 160%%;
                      %%1955, Soviet Phys. JETP 1, 128
\re
8.\ Kocharian N. M. et al.\ 1958, J. Exper. Theoret. Phys. (USSR) 35, 1335%%;
                      %%1959, Soviet Phys. JETP 35, 933
\re
9.\ Maeno T. et al.\ 2001, Astropart. Phys. 16, 121
\re
10.\ Motoki M. et al.\ 2003, Astropart. Phys. 19, 113
\re
11.\ Orito S.\ 1987, Proc. ASTROMAG Workshop, KEK Report KEK87-19, 111
\re
12.\ Orito S. et al.\ 2000, Phys. Rev. Lett. 84, 1078
\re
13.\ Sanuki T. et al.\ 2000, ApJ 545, 1135
\re
14.\ Sanuki T. et al.\ 2002, Phys. Lett. B 541, 234
\re
15.\ Sembroski G. H. et al.\ 1986, Phys. Rev. D 33, 639
\re
16.\ Shea M. A. and Smart D. F.\ 2001, Proc. 27th ICRC(Hamburg), 4063
\re
17.\ Shikaze Y. et al.\ 2000, Nucl. Instr. and Meth. A 455, 596
\re
18.\ Stephens S. A.\ 1997, Astropart. Phys. 6, 229

\endofpaper
\end{document}